\documentclass[aps,prb,twocolumn,superscriptaddress,floatfix,amsmath,amssymb,longbibliography]{revtex4-2}

\usepackage{graphicx}

\usepackage[colorlinks=true,citecolor=blue,linkcolor=blue,urlcolor=blue,bookmarks=false]{hyperref}
\usepackage{CJK}

\def\unit#1{\mathord{\thinspace\rm #1}}

\usepackage{chemformula} 
\usepackage[T1]{fontenc} 

\usepackage{times}
\usepackage{xcolor}
\usepackage{amsmath}
\usepackage{amssymb}
\usepackage{graphicx}
\usepackage{diagbox}
\usepackage{multirow}
\usepackage{threeparttable}
\usepackage{array}

\usepackage{mathtools}
\usepackage{braket}
\usepackage[makeroom]{cancel}
\graphicspath{ {images/} }

\begin{document}
\begin{CJK*}{UTF8}{bsmi}

\title{Fractal Quantum Transport on \ch{MoS2} Superlattices: a system with tunable symmetry.}

\author{Aitor Garcia-Ruiz (艾飛宇)}

\email{aitor.garcia-ruiz@phys.ncku.edu.tw}

\affiliation{Department of Physics and Center for Quantum Frontiers of Research and Technology (QFort), National Cheng Kung University, Tainan 70101, Taiwan}
\affiliation{Department of Physics and Astronomy, University of Manchester, Oxford Road, Manchester, M13 9PL, United Kingdom}
\affiliation{National Graphene Institute, University of Manchester, Booth St.\ E., Manchester, M13 9PL, United Kingdom}

\author{Ming-Hao Liu (劉明豪)}

\email{minghao.liu@phys.ncku.edu.tw}

\affiliation{Department of Physics and Center for Quantum Frontiers of Research and Technology (QFort), National Cheng Kung University, Tainan 70101, Taiwan}

\date{January 2024}

\begin{abstract}
Electron doping is an excellent tuning knob to explore different phases of matter in two-dimensional (2D) materials. For example, tuning the Fermi level at a van Hove singularity in twisted bilayer graphene can enhance electron-electron interactions and induce a diverse range of correlated phases \cite{cao_unconventional_2018,cao_correlated_2018}. Here, using a single-particle picture, we study the electronic reconstruction of the band edges of a 2D semiconductor, monolayer \ch{MoS2}, on a hexagonal moir\'{e} potential induced by another \ch{MoS2} monolayer. We find that such system transitions from a honeycomb to a hexagonal symmetry when the Fermi level is tuned from the conduction to the valence side. We also study the system under magnetic fields, and construct the Hofstadter's butterfly in the electron- and hole-doped side. Our findings are confirmed by simulating the conductance across a large-scale two-terminal device. We conclude that this duality is a general property that \ch{MoS2} and other transition-metal-dichalcogenides exhibit under non-symmetric superlattice potentials. 

\end{abstract}

\maketitle

\end{CJK*}

\section{Introduction}

Gate voltages are essential device components \cite{Anderson_Fundamentals_2004}. By controlling the electron concentration of semiconductors, they make transistors possible, which are at the cornerstone of today's technology \cite{zhang_review_2020}. From the point of view of material science, however, controlling the electron concentration extends beyond tuning the Fermi level. During the last decades, scientists have applied this concept to two-dimensional systems like composed of graphene or transition-metal-dichalcogenides (TMDs) to gain access to a rich variety of phases, including superconductivity and correlated insulators \cite{cao_unconventional_2018,cao_correlated_2018}, nematicity \cite{rubio-verdu_moire_2022}, charge density waves \cite{luckin_controlling_2024} or ferromagnetism \cite{seiler_quantum_2022,Zhou_Isospin_2022,de_la_barrera_cascade_2022}, to name just a few, where gate voltages can fully exploit their two-dimensional (2D) nature. Among this family of materials, \ch{MoS2}, a 2D semiconductor with a direct band gap of $\sim 1.8\unit{eV}$ at the K-point \cite{mak_atomically_2010,splendiani_emerging_2010}, has stood up as a candidate to lead the next generation of electronic devices \cite{radisavljevic_single-layer_2011,pham_mos2-based_2019,kumar_mos2-based_2020}, where gating enables us to control the electron concentration \cite{liao_mos2_2019} and external electric fields can even induce superconductivity \cite{taniguchi_electric-field-induced_2012}. More recently, with the advent of twistronics \cite{hennighausen_twistronics_2021}, engineering stacks of \ch{MoS2} monolayers with a relative angle between them has allowed to further extend the toolbox of methods for exploring new physics in \ch{MoS2} heterostructures \cite{weston_interfacial_2022}.

\begin{figure}[b]
\includegraphics[width=0.85\columnwidth]{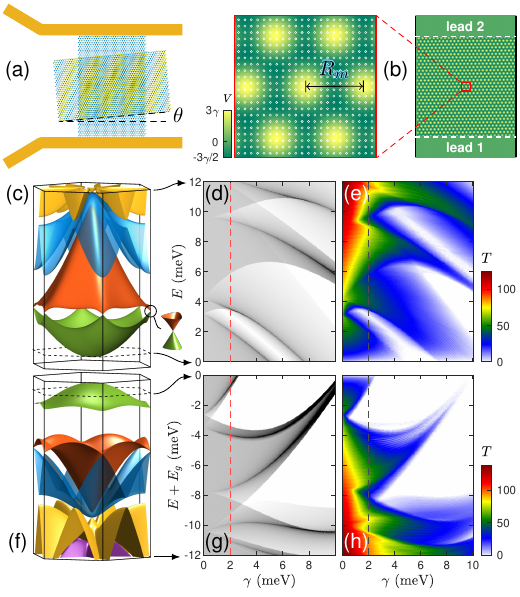}
\caption{(a) Schematic of a two-terminal \ch{MoS2} device stacked by a twisted, uncontacted \ch{MoS2} layer, which generates a superlattice potential $V(\mathbf{r})$ modeled by Eq.\ \eqref{Eq:V(r)}. (b) Colormap of $V(\mathbf{r})$, with white empty circles indicating the position grid points for the finite-difference-based square lattice for quantum transport. (c) Miniband structure close to the conduction band bottom, with $\gamma=2\unit{meV}$ and $R_m\approx 18.47\unit{nm}$ ($\theta= 1^\circ$). (d) DoS map and (e) conductance across the two-terminal device in (a) as a function of $\gamma$ and Fermi level. Panels (f)--(h) display a similar analysis for the valence band edge.\label{Fig:intro}} 
\label{Fig:zeroBfield}
\end{figure}

On the other hand, applying magnetic fields has also been used to study the properties of \ch{MoS2} \cite{goryca_revealing_2019,pisoni_interactions_2018}. Typically, weak magnetic fields ($B\sim 1$--$10\unit{T}$) discretizes the electronic spectrum in a series of Landau levels (LL), which generates distinctive signatures in the quantum transport, such as Shubnikov-de Haas oscillations \cite{cui_multi-terminal_2015,kormanyos_landau_2015,smolenski_interaction-induced_2019}. For values of the magnetic length comparable to those of the interactomic distances of ($\sim1$ \AA), it is expected that the LL spectrum of \ch{MoS2} undergo a strong reconstruction, evolving into a series of complex self-similar structures \cite{ho_hofstadter_2015}, also known as \emph{Hofstadter's butterfly} \cite{hofstadter_energy_1976}. However, the values of the magnetic field required for its observation are orders of magnitude greater than those achievable in experiments. In this context, superlattice potentials can introduce a much larger nano-meter length scale in the system, thus enabling the observation of the Hoftadter's spectrum in 2D materials for much lower values of the magnetic fields. This concept has been brought to graphene superlattices, where it has been extensively studied both theoretically \cite{bistritzer_moire_2011, hejazi_landau_2019, crosse_hofstadter_2020,moon_energy_2012, lian_open_2021, fabian_half-integer_2022,chen_dirac_2014,ponomarenko_cloning_2013} and experimentally \cite{huber_gate-tunable_2020,krishna_kumar_high-temperature_2017,krishna_kumar_high-order_2018,barrier_long-range_2020, huber_band_2022,yang_hofstadter_2016}. However, a similar study on \ch{MoS2} superlattices, as well as an analysis of its quantum transport, is still missing.

In this work, we investigate the electronic structure and the conductivity, with and without a perpendicular magnetic field, of a monolayer \ch{MoS2} under a hexagonal superlattice potential. We find that the conduction and valence edges reconstruct following a fundamentally different symmetry, namely, honeycomb and hexagonal, respectively. To demonstrate this, in Sec. \ref{Sec:2}, we describe our methodology to compute the electronic spectrum and the conductance across the two-terminal device depicted in \autoref{Fig:zeroBfield} (a). We study such system under magnetic fields in Sec. \ref{Sec:3}. The dispersion shows two different Hofstadter's spectra for the conduction and valence band, which for the lowest bands resembles the Hofstadter's butterfly of graphene and a hexagonal lattice, respectively. The conductance, in turn, features two distinct regimes of quantum transport, bulk and edge transport, and provide an interpretation in terms of the topological nature of the bands. Our final remarks are provided in Sec. \ref{Sec:4}.

\section{Miniband spectrum and quantum transport}\label{Sec:2}

Our work focuses on the energy range close to the conduction and valence band edges, located at the corners of the Brillouin zone, where \ch{MoS2} hosts a direct band gap of about $\sim1.8$ eV \cite{kadantsev_electronic_2012}. At these points, the character of the wavefunction at the conduction and valence side is mostly prescribed by the $d_z$ and $d_{xy}+d_{x^2-y^2}$ orbitals of molybdenum, respectively, which form a hexagonal lattice with lattice constant $a\approx 3.18 \unit{\text{\AA}}$ \cite{liu_three-band_2013,fang_ab_2015,shahriari_band_2018,kang_band_2013}. Around the K-point, therefore, the spectrum of \ch{MoS2} is effectively described as two decoupled two-dimensional electron gas (2DEG), with masses $m_e\approx0.46 m_0$ and $m_h\approx-0.55 m_0$ ($m_0$ is the free electron mass), in the conduction and valence side, respectively. 

When an additional TMD monolayer is stacked on top of \ch{MoS2} with a small misalignment angle $\theta$, a moir\'{e} potential emerges with a much larger lattice constant, which for homobilayers, is given by $\mathbf{R}_{m}\!\!=\!\!{a}/{2\sin(\theta/2)}$.
The system described above has been extensively analysed as a joint structure of two equally relevant monolayers of TMD, due to their potential to engineer topological insulators, flat bands or skyrmion lattice textures \cite{devakul_magic_2021,wu_topological_2019,kundu_moire_2022,zhang_electronic_2021,tang_geometric_2021}. Here, in contrast, we study the electrons inside one monolayer of \ch{MoS2} affected by a generic hexagonal superlattice potential, which originates from its proximity to another TMD monolayer, as shown in Fig. \ref{Fig:intro} (a). In particular, we consider that the additional monolayer is also \ch{MoS2}, and that it forms an angle of $\theta=1^\circ$, which induces by proximity a moir\'{e} potential with lattice constant $R_m\approx18.47 \unit{nm}$. The Hamiltonian for one electron in such system takes the form
\begin{equation}\label{Eq:H(r)}
    \mathcal{H}_{\beta} = \frac{-\hbar^2}{2{m}_{\beta}}\nabla^2+V(\mathbf{r})\ ,
\end{equation}
where $\beta=e,h$ is the band index, and
\begin{equation}\label{Eq:V(r)}
V(\mathbf{r}) =
\gamma\sum_{j=1}^3
\cos(\mathbf{G}_j\cdot\mathbf{r})
\end{equation}
is the model potential, expanded up the smallest Fourier harmonics $\mathbf{G}_j=G\left[\sin\left({j2\pi}/{3}\right),-\cos\left(j{2\pi}/{3}\right)\right]$ with $G={4\pi}/{\sqrt{3}R_m}$. Note that, despite the simplicity of Eq.\ \eqref{Eq:V(r)}, such potential was first adopted to describe the superlattice in graphene due to the aligned hexagonal boron nitride (hBN) lattice \cite{yankowitz_emergence_2012}, and later found to give satisfactory agreement in quantum transport simulations choosing $\gamma\approx 6\unit{meV}$ \cite{chen_electrostatic_2020,kraft_anomalous_2020}. In our case, the exact value for the indirect coupling between d-orbitals of \ch{MoS2} is unknown, although several \textit{ab initio} studies suggest that such coupling is generally weak, of the order of $\sim1$ meV \cite{xian_realization_2021,liu_evolution_2014,chang_thickness_2014,zhong_interfacial_2016,liu_tuning_2012,huang_probing_2014}. Therefore, we will take the value $\gamma = 2$ meV and apply the continuum model \cite{lopesdossantos_graphene_2007,bistritzer_moire_2011-1,koshino_interlayer_2015,garcia-ruiz_full_2021} to diagonalise the Hamiltonian in Eq. (\ref{Eq:H(r)}), where the first term is diagonal in the basis of plane waves  $\psi_{\mathbf{k}}=\exp(i\mathbf{k}\cdot\mathbf{r})/N$, -- $\mathbf{k}=(k_x,k_y)$ and $N$ is a normalisation factor -- and the second term couples plane waves that differing by one reciprocal superlattice vector $\pm\mathbf{G}_j$. Expanding the basis of plane waves $\Psi_{\mathbf{k}}=\{\psi_{\mathbf{k}},\psi_{\mathbf{k}-\mathbf{G}^1},\psi_{\mathbf{k}-\mathbf{G}^2},...\}$, the Hamiltonian takes the matrix form
\begin{equation}
\mathcal{H}_{\mathbf{k}}^{\beta}=
\begin{pmatrix}
\frac{\hbar|\mathbf{k}|^2}{2{m}_{\beta}} & \frac{\gamma}{2} & \frac{\gamma}{2} & \cdots\\
\frac{\gamma}{2} & \frac{\hbar|\mathbf{k}-\mathbf{G}_1|^2}{2{m}_{\beta}} & 0 & \cdots\\
\frac{\gamma}{2} & 0 & \frac{\hbar|\mathbf{k}-\mathbf{G}_2|^2}{2{m}_{\beta}} & \cdots\\
\vdots & \vdots & \vdots & \ddots
\end{pmatrix}\ ,
\label{Eq:HamMatrix}
\end{equation}
which becomes finite after truncating the basis $\Psi_{\mathbf{k}}$ to some maximum value of wavenumber shift $\mathbf{G}<k_c$ that ensures convergence of the miniband spectrum. From the band structure $\epsilon_{\mathbf{k}}$, the density of states (DoS) can be computed numerically using,
\begin{align}\label{Eq:DoS}
    \rho(E)=g
    \int \frac{d\mathbf{k}}{(2\pi)^2}\delta(E-\epsilon_{\mathbf{k}}),
\end{align}
where $g$ is the degeneracy of the bands. In graphene systems, it is usual to take $g=4$, reflecting the valley and spin degeneracy. In \ch{MoS2}, however, due to the spin-orbit coupling, the conduction and valence bands are split into two spin up and spin down polarized bands by about $15$ meV \cite{pisoni_interactions_2018} and $170$ meV \cite{latzke_electronic_2015}, respectively. In this work, we will consider an energy range of 12 meV from the band edges, which effectively gives a band edge degeneracy of $g=2$. To note, the range of electron doping corresponding to the filling of our range of energies is $n\approx2.32\times10^{12}\mathrm{cm}^{-2}$ ($n\approx-2.83\times10^{12}\mathrm{cm}^{-2}$) in the conduction  (valence) band.

In \autoref{Fig:zeroBfield} (c), we show the miniband spectrum for $\gamma=2$ meV, with the Fermi level tuned into the conduction side. The lowest two minibands exhibit a graphene-like dispersion, with touching points at the corners of the mini-Brillouin zone (mBZ), mimicking the low-energy Dirac-like dispersion of graphene. This resemblance originates from the position at which the low-energy states are localized in real space, around the minima the model potential $V(\mathbf{r})$ in Fig. \ref{Fig:zeroBfield} (b),  forming a honeycomb lattice (see supplementary material). This feature is known to emerge in generic two-dimensional electron gases (2DEG) under hexagonal potentials, which is why it is often referred to as \textit{artificial graphene} \cite{polini_artificial_2013, krix_artificial_2020, chen_artificial_2021}. On panel (d) of the same figure, we present the DoS for a broad range of $\gamma$, where the red vertical dashed line marks the case $\gamma=2$ meV shown in Fig. \ref{Fig:zeroBfield} (c). From the DoS map, we observe that, as $\gamma$ becomes larger, the first two bands become completely isolated, and the third band flattens, spanning about 1 meV. 

In panel (f) of \autoref{Fig:zeroBfield}, we present the miniband structure for hole doping. In contrast to the panel above, here the miniband closest to the charge neutrality point (CNP) is spectrally isolated from the rest, and resembles the dispersion obtained from a tight-binding model on a hexagonal lattice. In contrast to the conduction side, states of the highest miniband are localized around the yellow maxima in Fig. \ref{Fig:intro} (b), which form a hexagonal lattice. Thus, in contrast to the conduction side, the first miniband of the hole-dope spectrum is reconstructed following an \textit{artificial hexagonal lattice}. 

In a transport experiment, it is expected that the two different lattice reconstructions for electron and hole doping would reproduce two different responses. To demonstrate this, here we also simulate the conductance across a realistic two-terminal device, sketched on the right hand side of \autoref{Fig:zeroBfield} (b). To this aim, we follow the real-space Green's function method \cite{datta_electronic_1995,buttiker_four-terminal_1986}, where the band edges of \ch{MoS2} are modelled using a tight-binding model on a square lattice [empty white circles in panel (b)] over a square section of $500\times500$ nm, where the Hamiltonian takes the form
\begin{equation}\label{Eq:H(r)_TB}
\mathcal{H}= 
\sum_{\beta=e,h}
\left[
t_\beta
\sum_{\langle \mathrm{i},\mathrm{j}\rangle}
\hat{c}_{\beta,\mathrm{i}}^\dag
\hat{c}_{\beta,\mathrm{j}}+
\sum_n
V_n
\hat{c}_{\beta,n}^\dag
\hat{c}_{\beta,n}
\right].
\end{equation}
Above, $\hat{c}_{\beta,\mathrm{i}}$ ($\hat{c}_{\beta,\mathrm{i}}^{\dag}$) is the operator that annihilates (creates) an electron in the band $\beta=e,h$ at the lattice site $\mathrm{i}$, $\langle \cdots \rangle$ denotes nearest neighbours and $t_\beta=-\hbar^2/2m_\beta a^2$ is the hopping parameter, which reproduces the same parabolic dispersion at the band edges as the Hamiltonian in Eq. (\ref{Eq:H(r)}). The on-site energy $V_n=V(\boldsymbol{r}_n)$ captures the effect of the model potential in Eq. (\ref{Eq:V(r)}) within the scattering region. This region is connected to two semi-infinite leads, across which the conductivity is computed using $\mathcal{G}=2(e^2/h)T$, with $T$ being the transmission probability, that is, the sum over all propagating modes across the two leads \cite{imry_conductance_1999}. 

In \autoref{Fig:zeroBfield} (e) and (h), we present the transmission in the conduction and valence side, respectively, for values of the superlattice strength $\gamma$ ranging 0 to 10 meV. As expected, the numerically computed transmission is consistent with the miniband spectrum constructed using the continuum model: low density states in the scattering regions correlates with low transmission. This evidences that transmission also reflects the symmetry duality for electrons and holes.

\section{Hofstadter's butterfly and fractal transport}\label{Sec:3}

In this section, we analyse the changes in the dispersion and the transport induced by an external magnetic field $\mathbf{B}=(0,0,B)$. The first term in Eq.\ \eqref{Eq:H(r)} generates a discrete spectrum of Landau levels (LL) $E_n=(n+1/2)\hbar\omega$, with $n\in\mathbb{Z}_{\geq0}$, and $\omega=eB/m_\beta$. The eigenstates associated with these LLs are coupled to each other by the second term. In particular, using the Landau gauge $\mathbf{A}=(-yB,0,0)$, eigenstates with guiding centres differing by an amount $\Delta$ are coupled. As explained in the supplementary material, the value of this distance is fixed by $\frac{p}{q}=\frac{h/e}{2\Phi}$, where $p,q$ are coprimes integers and $\Phi$ is the magnetic flux across the non-magnetic unit cell. We obtain the matrix elements of the Hamiltonian using a larger unit cell composed of magnetic Bloch functions,
\begin{align}
|n,y_0,j,k_y\rangle=\frac{1}{\sqrt{M-1}}\sum_{m=0}^M
    e^{i(mq+j)\Delta k_y}|n,y_0+(mq-j)\Delta\rangle.
\end{align}
Above, $y_0$ is the guiding centre, $M$ is the number of magnetic unit cells, $q$ is the number of Landau levels within the magnetic unit cell, and $j$ is an integer that goes from $0$ to $q-1$ (see Supplementary Material for further details). The eigenvalues of the resulting Hamiltonian are used to compute the band structure while the eigenvectors encode the information to obtain the Chern numbers associated to each band, which we compute following the references \cite{fukui_chern_2005, hejazi_landau_2019}. In the transport calculations, we incorporate the magnetic field in the scattering region by the Peierls substitution \cite{peierls_zur_1933}, where the hopping parameter acquires a space-dependent complex phase factor $t_{\beta}\to t_{\beta}\exp(i\frac{e}{\hbar}\int_\mathrm{i}^\mathrm{j} \mathbf{A\cdot d\mathbf{r}})$. Note that in order to compute a magnetic band structure, only a discrete set of magnetic fields constrained by the commensurate condition are allowed, whereas in transport simulations one can tune continuously the value for the magnetic field strength. 

\begin{figure}
\begin{center}
\includegraphics[width=1\columnwidth]{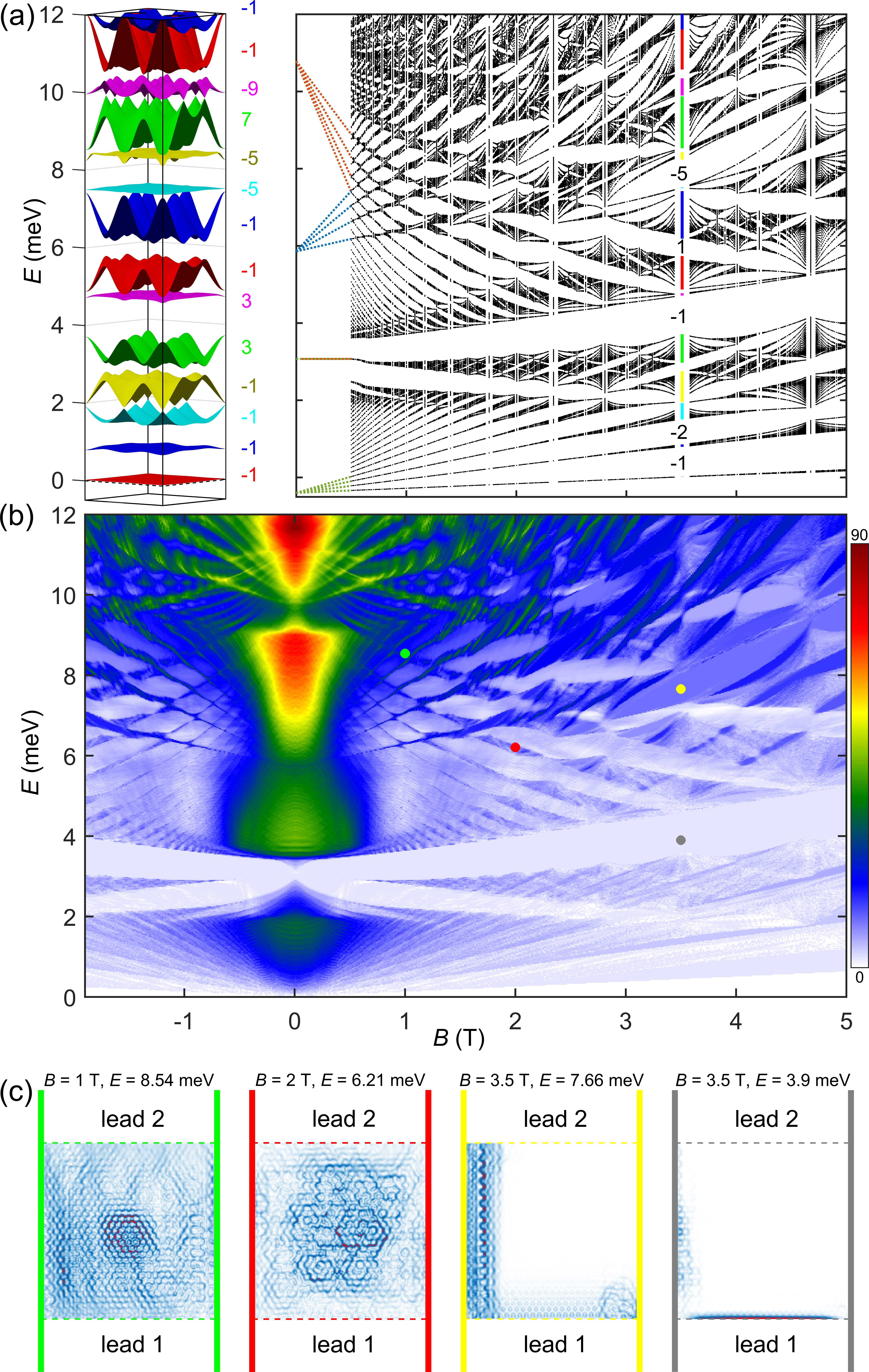}
\caption{ (a) Hofstadter's spectrum of the conduction band of \ch{MoS2} in a triangular superlattice potential with lattice constant $18.47$ nm and coupling strength $2$ meV. On the right, we present an exemplary band structure for the commensurate structure  $p=2,q=1$ ($B\approx3.5$ T) (b) Transmission across a two-terminal device. (c) Current density plots for the four marks in the panel above, to highlight the bulk transport and the edge transport. To mention, the numerically computed transport for the rightmost panel, with Chern number -5, is exactly 5 times larger than that of the middle panel, with Chern number -1. \label{Fig:nonzeroBfield_conduction}}
\end{center}
\end{figure}

\begin{figure}
\begin{center}
\includegraphics[width=1\columnwidth]{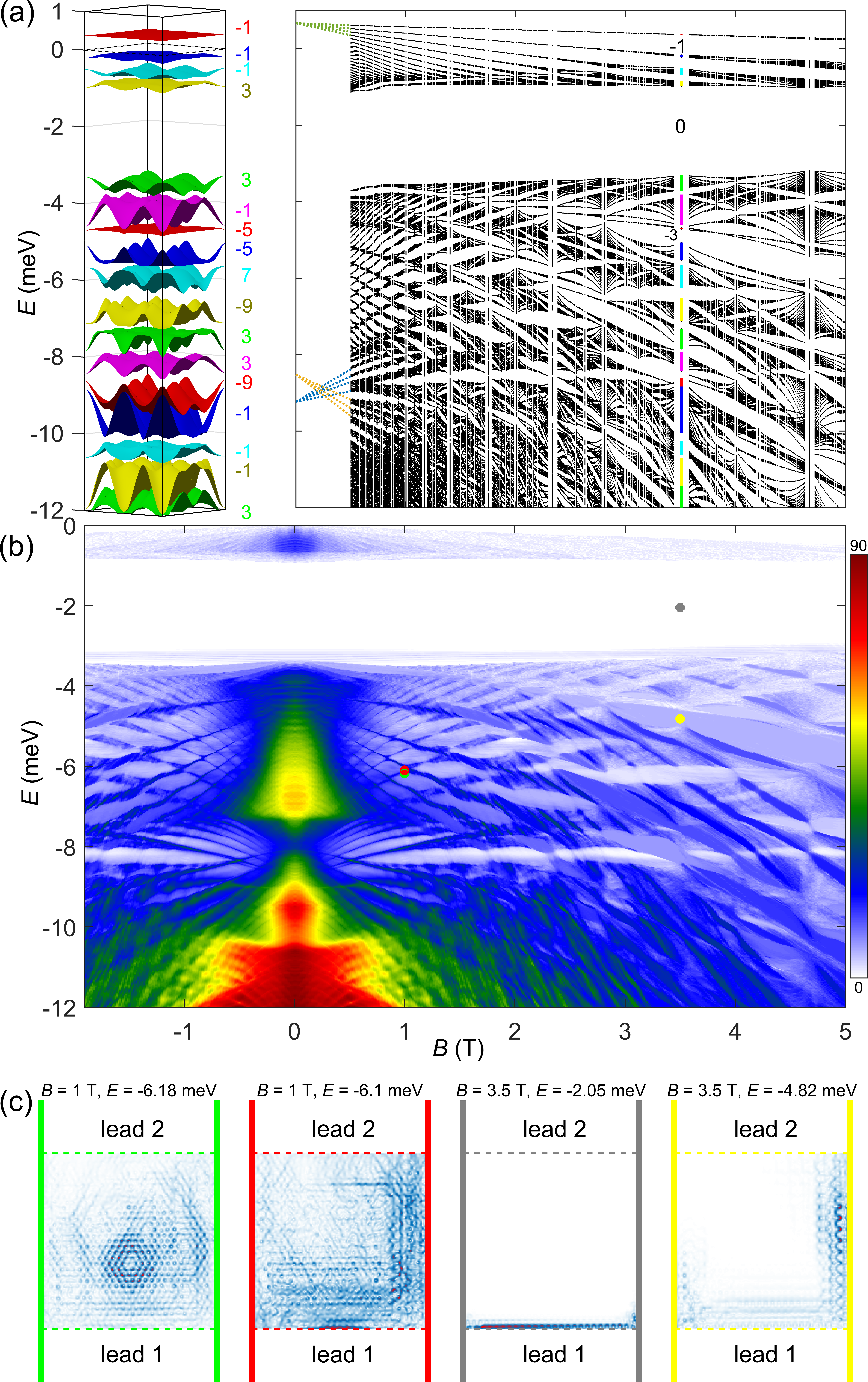}
\caption{ (a) Hofstadter's butterfly of the valence band of \ch{MoS2} in a hexagonal superlattice potential with lattice constant $18.47$ nm and coupling strength $2$ meV. On the right, the band structure and the associated Chern numbers for the commensurate structure $p=2, q=1$ ($B\approx3.5$ T) (b) Transmission across a two-terminal device. (c) Current density plots for the three marks in the panel above, showing bulk transport at a Landau level, no transport inside topologically trivial gaps and edge transport. \label{Fig:nonzeroBfield_valence}}
\end{center}
\end{figure}

Magnetic fields induces strong reconstructions in the mini band structure of the conduction band edge described Sec. \ref{Sec:2}. For $B\lesssim0.5$, the spectrum in the conduction band turns into a series of Landau levels, linear in magnetic field, and with origins at the parabolic band edges of the non-magnetic bands, as shown in dotted lines in  Fig. \ref{Fig:nonzeroBfield_conduction} (a). For magnetic fields $\lesssim1$ T, we observe a rhomboid mesh-like structure at $\sim8$ meV, originated from the inter-crossing of Landau levels coming from the top edge of the second and the bottom edge of the third non-magnetic mini bands. At higher magnetic fields, these rhomboid mesh develops into an intricate self-similar structure of larva-like energy windows of absence of states, which co-exists with other butterfly-like structures. We also observe two windows of empty states, separated by a fractal structure that originates from $\sim3$ meV, which corresponds to the zeroth order Landau level of the conical dispersion in Fig. \ref{Fig:zeroBfield} (b), for very low magnetic fields.

The quantum transport simulation across a two-terminal device is shown in Fig. \ref{Fig:nonzeroBfield_conduction} (b). Overall, the transport map exhibits identical features to those in the Hofstadter's butterfly computations, including the rhombiod mesh-like structures and the larva-shaped features. However, unlike in Figs. \ref{Fig:zeroBfield} (b) and (c), transport inside band gaps is not always zero, but $g$ times an integer value of the fundamental conductance, $e^2/h$. This is because gaps host topologically protected edge states, where the transport takes place across the edges of the sample, and we confirm by computing the Chern numbers associated to each magnetic mini band. For example, inside each gap of the band structure that corresponds to $p=2,q=1$ ($B\approx3.5$ T), the numerically computed value for the conductance is equal to the sum of the Chern numbers of all the bands below [shown inside the Hofstadter's map in Fig. \ref{Fig:nonzeroBfield_conduction} (a)] times $ge^2/h$. Conversely, quantum transport within the bands occurs inside the bulk. To visualise these two types of quantum transport, we present the current density amplitude plot in panel (c) of the same figure. While in bulk transport the maximum amplitude of the current density is at the centre of the sample, in edge transport, the centre of the sample features absence of states.

When the Fermi level is tuned in the valence side, the electronic properties of the system are prescribed by a totally different symmetry order, a hexagonal lattice. In \autoref{Fig:nonzeroBfield_valence} (a), we extend our previous study on the band reconstruction to the valence band edge of \ch{MoS2}. While the electronic spectrum have some commonalities with its conduction counterpart, such as a rhombiod-like structure at about $B\sim1$ T, the Hofstadter's butterfly in the valence side is totally different, with a topologically trivial gap from $-3$ to $-1$ meV, where the sum of all Chern numbers above the energy gap is zero. In panel (b) of the same figure, we show that these features are in agreement with our quantum transport simulations. In particular, we observe bulk transport at the Landau levels, zero conductance inside the trivial gap, and quantized conductance inside topological gaps [see panel(c)].

\section{Conclusions}\label{Sec:4}

In this work, we study the changes in the band structure and the quantum transport of a 2D semiconductor, \ch{MoS2}, under the effect of a hexagonal superlattice potential. We demonstrate that the conduction and valence band edges reconstruct following different symmetries: while states in the former reconstruct following a honerycomb superlattice, the band structure changes in latter are prescribed by a hexagonal superlattice. We also investigate its electronic spectrum under magnetic fields, which inherits such duality in the form of two different Hofstadter's butterflies for the conduction and valence bands. These features are confirmed by simulating the quantum transport across a two-terminal device, where we also identify two distinct regimes of quantum transport, namely, bulk transport, which characterises the conductance inside bands, and edge transport, where the material conducts due to the non-trivial topology of the gaps. 

Finally, we have proposed twisted \ch{MoS2} bilayers as an experimental platform to realized the above-described findings, due to the tunability that van der Waals heterostructures offer. However this concept can be applied to a broad range of semiconductors, like Galium Arsenide \cite{gibertini_engineering_2009, de_simoni_delocalized-localized_2010, singha_two-dimensional_2011, nadvornik_laterally_2012, goswami_transport_2012}, on a nanopatterned periodic potential where their maxima and minima form different lattice symmetries, such as Lieb or Kagom\'{e} lattices.

\section{Acknowledgements}\label{Sec:5}

We thank Prof. M. Mucha-Kruczynski for useful discussions. We also thank National Science and Technology Council (NSTC 112-2112-M-006-019-MY3) for financial support and National Center for High-performance Computing (NCHC) for providing computational and storage resources.

\bibliographystyle{unsrt}
\bibliography{Bibl}

\end{document}


\begin{CJK*}{UTF8}{bsmi}

\title{Supplementary material of $``$Fractal Quantum Transport on \ch{MoS2} Superlattices: a system with tunable symmetry.$"$}

\author{Aitor Garcia-Ruiz (艾飛宇)}

\affiliation{Department of Physics and Center for Quantum Frontiers of Research and Technology (QFort), National Cheng Kung University, Tainan 70101, Taiwan}
\affiliation{Department of Physics and Astronomy, University of Manchester, Oxford Road, Manchester, M13 9PL, United Kingdom}
\affiliation{National Graphene Institute, University of Manchester, Booth St.\ E., Manchester, M13 9PL, United Kingdom}

\author{Ming-Hao Liu (劉明豪)}

\affiliation{Department of Physics and Center for Quantum Frontiers of Research and Technology (QFort), National Cheng Kung University, Tainan 70101, Taiwan}

\date{January 2024}
\maketitle

\end{CJK*}

\section{Spatial localization of wavefunctions in the first minibands.}

\begin{figure}
\begin{center}
\includegraphics[width=1\columnwidth]{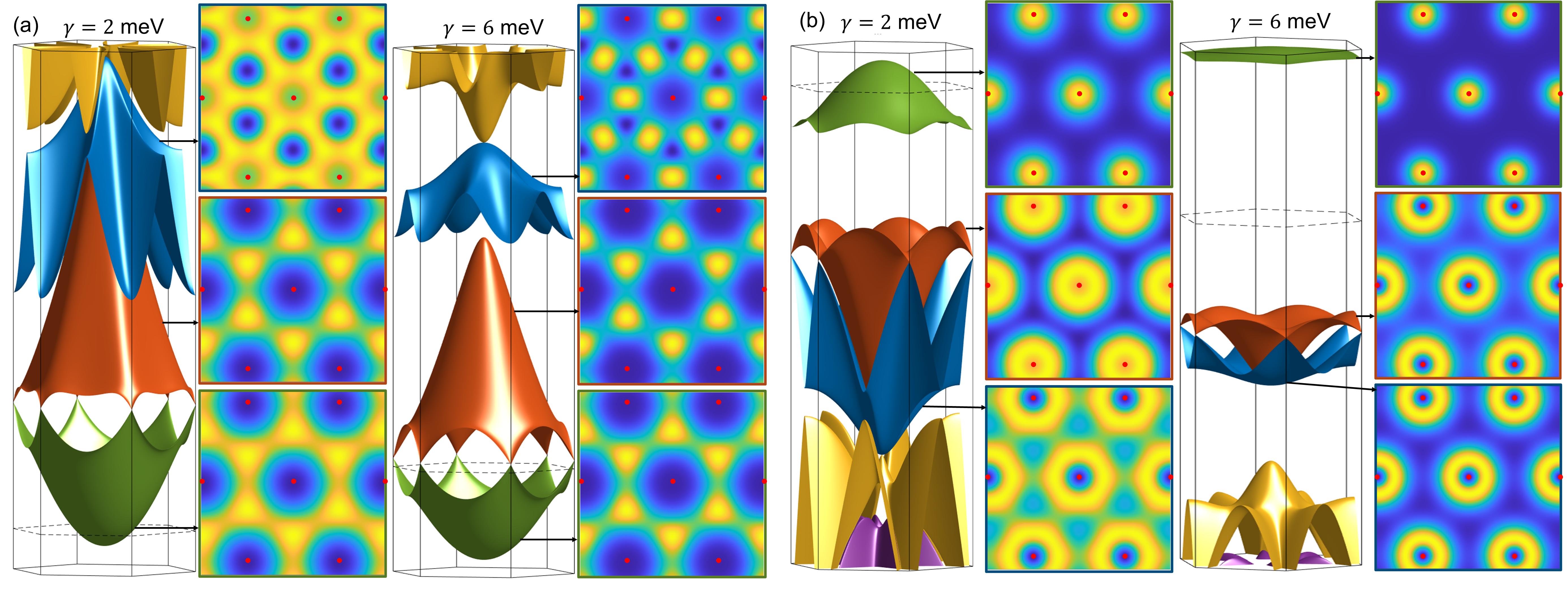}
\caption{ (a) Mini band structure of the conduction band edge and band-integrated charge densities, $n_e^{1,2,3}(\mathbf{r})$ across the $\mathrm{MoS}_2$ monolayer for $\gamma =$ 2 meV and 6 meV. We also represented the maxima of $V(\mathbf{r})$ with red dots to highlight that the first two bands are localized around the minima of the superlattice potential, and this localization increases with $\gamma$. (b) Same as in (a) in the valence side, where $n_h^{1}(\mathbf{r})$ features localization around the maxima of $V(\mathbf{r})$. \label{Fig:Localization}}
\end{center}
\end{figure}

In this section, we present the spatial charge distribution of states for the lowest two (highest) mini bands in the conduction (valence) side of $\mathrm{MoS}_2$ under the triangular superlattice potential $V(\mathbf{r})$ in Eq. (1) of the main text, without a magnetic field. Following previous works \cite{garcia-ruiz_electronic_2020}, we compute the squared amplitude of the eigenvector
\begin{align}
    \Phi_{\beta}^{b}(\mathbf{k},\mathbf{r})=
    \frac{1}{\sqrt{N}}
    \sum_{\mathbf{G}}
    c_{\beta\mathbf{G}}^{b}(\mathbf{k})
        e^{i(\mathbf{k}-\mathbf{G})\cdot\mathbf{r}},
\end{align}
where the the term $c_{\beta\mathbf{G}}^{n}(\mathbf{k})$ refers to the component of the wavefunction associated to the Bloch function shifted by $\mathbf{G}=a_1\mathbf{G}_1+a_2\mathbf{G}_2$, in the $b$-th band in the valence ($\beta=h$) or conduction ($\beta=e$) side at the point $\mathbf{k}$. Here, in order to demonstrate the localization around the minima (maxima) of the potential profile shown in Fig. 1 (a) of the main text of the first two (one) bands in the conduction (valence) side of our system, we integrate the squared amplitude of the expression above over the first mini Brillouin zone, 
\begin{align}
    n^b_\beta(\mathbf{r})=
    g
    \int_{\mathrm{mBZ}}
    \frac{d\mathbf{k}}{(2\pi)^2}
    \left\vert
    \sum_{\mathbf{G}}
    c_{\beta\mathbf{G}}^{b}(\mathbf{k})
    e^{-i\mathbf{G}\cdot\mathbf{r}}
    \right\vert^2
\end{align}

In Fig. \ref{Fig:Localization}, we plot $n^{b}_{e/h}$ for the three minibands closest to the band gap, using two values of the superlattice coupling parameter, $\gamma=2$ meV and $\gamma=6$ meV. In the conduction side, our color maps demonstrates that the first two minibands are indeed localized around the minima of $V(\mathbf{r})$, which form a honeycomb lattice, and this localization increases with $\gamma$. In turn, the third band features hot spots of electronic amplitude in the mid-points of these minima, which form a a Kagome-type of lattice for large values of $\gamma$. On the valence side, the highest energy states are localized at the lattice points of a hexagonal lattice. As the parameter $\gamma$ gets stronger, the electronic distribution becomes more localized and the topmost band turns flatter and flatter.

\section{Magnetic band structure and Hofstadter's butterfly spectrum.}

In this section, we expand on the method used to compute the Hofstadter's butterfly and the mini-band spectrum shown in Figs. (2) and (3) of the main text. The spectrum of the valence and conduction band edges of $\mathrm{MoS}_2$ under the effects of a magnetic field, turns into a series of perfectly flat bands, known as Landau levels \cite{Landau_Diamagnetismus_1930}. Using the Landau gauge $\mathbf{A}=(-yB,0,0)$, the eigenvalues and eigenvectors take the form
\begin{subequations}\label{Eq:LL_basis}
\begin{align}
E_{\beta,n}=&
(n+1/2)\hbar\omega_\beta,\qquad  \omega_\beta=\frac{eB}{m_\beta},\\
|n,y_0\rangle=&
\frac{1}{\sqrt{L_x}}
e^{ik_xx}\cdot
\psi_n\left(\frac{y-y_0}{l_B}\right),\qquad l_B=\sqrt{\frac{\hbar}{m\vert\omega_\beta\vert}},
\qquad
y_0=k_x l_B^2
\end{align}
\end{subequations}
where $\psi_n(x)=\left(2^n n!\sqrt{\pi}\right)^{-1/2}H_n(x)\,
e^{- \frac{x^2}{2}}$, and $H_n(x)$ are the Hermite polynomials \cite{Olver_NIST_2010}. As shown below, the periodic potential can couple two eigenvectors with guiding centres differing by a to-be-determined amount $\Delta$. We use the basis of unperturbed Landau levels in Eqs. (\ref{Eq:LL_basis}) to construct a larger (magnetic) unit cell by grouping together $q$ number of eigenvectors,
\begin{align}
    \{|n,y_0\rangle,
    |n,y_0+\Delta\rangle,
    |n,y_0+2\Delta\rangle,
    \cdots,|n,y_0+(q-1)\Delta\rangle\},
\end{align}
with $|n,y_0+\Delta q\rangle$ being equivalent to $|n,y_0\rangle$, and the number $q\in\mathbb{N}$ is the number of Landau levels in the magnetic unit cell. Assuming our sample contains $M$ unit cells, we introduce the magnetic Bloch functions,
\begin{align}\label{Eq:MagneticBlochFunction}
|n,y_0,j,k_y\rangle=\frac{1}{\sqrt{M}}\sum_{m=0}^M
e^{i(mq+j)\Delta k_y}|n,y_0+(mq+j)\Delta\rangle,
\end{align}
and use them to evaluate the matrix elements of the Hamiltonian, 
\begin{align}
\langle n',y_0',j',k'_y \vert
\mathcal{H}
\vert n,y_0,j,k_y \rangle=
E_{n}^\beta
\delta_{n',n}
\delta_{y_0',y_0}
\delta_{j',j}
\delta_{k_y',k_y}+
\frac{\gamma}{2}
\sum_{j=1}^3 
\langle n',y_0',j',k'_y \vert
e^{i\mathbf{G}_j\cdot\mathbf{r}}+
e^{-i\mathbf{G}_j\cdot\mathbf{r}}
\vert n,y_0,j,k_y \rangle.
\end{align}
The integration over the spatial coordinate $x$ in the second term of the expression above fixes the value for $\Delta$, while integration over the spatial coordinate $y$ gives the values for the magnetic field that ensures a commensurate structure. To illustrate this, we compute explicitly the contribution to the matrix elements from the first exponential term,
\begin{align}\label{Eq:ME_G1}
&\langle
n',y_0',j',k_y'
|
e^{i\mathbf{G}_1\cdot \mathbf{r}}
|
n,y_0,j,k_y\rangle=\langle
n',y_0',j',k_y'
|
e^{i\frac{\sqrt{3}}{2}Gx}
e^{i\frac{1}{2}Gy}
|
n,y_0,j,k_y\rangle\\
&=
\frac{1}{M}
\sum_{m',m}
e^{-i(m'q+j')\Delta k_y}
e^{i(mq+j)\Delta k_y}
\int_{-\infty}^\infty dx
\frac{
e^{i\left(\frac{\sqrt{3}}{2}G+k_x-k_x'\right)x}}{L_x}
\nonumber\\
&\times
\int_{-\infty}^\infty dy
e^{i\frac{1}{2}Gy}
\psi_{n'}
\left(\frac{y-y_0-(m'q+j')\Delta}{l_B}\right)
\psi_n
\left(\frac{y-y_0-(mq+j)\Delta}{l_B}\right)
\nonumber\\
&=
\frac{1}{M}
\sum_{m',m}
e^{-i[(m'-m)q+j'-j]\Delta k_y}
\delta_{k_x',k_x+\frac{\sqrt{3}G}{2}}
e^{i\frac{1}{2}Gy_0}
e^{i\frac{1}{2}Gq\Delta m}
e^{i\frac{1}{2}Gj\Delta}
\nonumber
\\
&\times
\int_{-\infty}^\infty d\tilde{y}
e^{i\frac{1}{2}G\tilde{y}}
\psi_{n'}
\left(\frac{\tilde{y}-[(m'-m)q+j'-j]\Delta}{l_B}\right)
\psi_n
\left(\frac{\tilde{y}}{l_B}\right),
\nonumber
\end{align}
where the Kronecker delta gives the value for the change in guiding centre, $\Delta=\frac{\sqrt{3}}{2}Gl_B^2$, and the complex exponential term, together with summation over $m$ and $m'$, cancels out unless $\frac{1}{2}Gq\Delta $ is a integer multiple of $2\pi$. Using relations in Eq. (\ref{Eq:LL_basis}) we obtain the condition for a commensurable structure
\begin{align}
\frac{p}{q}=
\frac{G\Delta}{4 \pi}=
\frac{h/e}{\sqrt{3}R_m^2 \times B}\equiv
\frac{\Phi_0}{2\Phi}.
\end{align}
Above, $\Phi_0$ is the fundamental unit of magnetic flux and $\Phi$ is the magnetic flux across one unit cell. Changing the variables $\tilde{y}=\bar{y}+\Delta/2$, Eq. (\ref{Eq:ME_G1}) reduces to
\begin{align}\label{Eq:ME_G1}
\langle
n',y_0',j',k_y'
|
e^{i\mathbf{G}_1\cdot \mathbf{r}}
|
n,y_0,j,k_y\rangle=&
\frac{1}{M}
\sum_{m',m}
e^{-i\Delta k_y}
e^{i\frac{1}{2}Gy_0}
e^{i\frac{1}{2}Gj\Delta}
e^{i\frac{1}{4}G\Delta}
\delta_{y_0',y_0}
\delta_{m'q+j',mq+j+1}
\\
&\times\int_{-\infty}^\infty d\bar{y}
e^{i\frac{1}{2}G\bar{y}}
\psi_{n'}
\left(\frac{\bar{y}-\Delta/2}{l_B}\right)
\psi_n
\left(\frac{\bar{y}+\Delta/2}{l_B}\right)
\equiv\hat{\mathcal{T}}_{j}^{(1)}
\nonumber
\end{align}
We proceed similarly with the contributions from the other two reciprocal lattice vectors, 
\begin{align}\label{Eq:ME_G2}
\langle
n',y_0',j',k_y'
|
e^{i\mathbf{G}_2\cdot \mathbf{r}}
|
n,y_0,j,k_y\rangle=&
\frac{1}{M}
\sum_{m',m}
e^{i\Delta k_y}
e^{i\frac{1}{2}Gy_0}
e^{i\frac{1}{2}Gj\Delta}
e^{-i\frac{1}{4}G\Delta}
\delta_{y_0',y_0}
\delta_{m'q+j',mq+j+1}\\
&\times\int_{-\infty}^\infty d\bar{y}
e^{i\frac{1}{2}G\bar{y}}
\psi_{n'}
\left(\frac{\bar{y}+\Delta/2}{l_B}\right)
\psi_n
\left(\frac{\bar{y}-\Delta/2}{l_B}\right)
\equiv\hat{\mathcal{T}}_{j}^{(2)}
\nonumber
\end{align}
\begin{align}\label{Eq:ME_G3}
\langle
n',y_0',j',k_y'
|
e^{i\mathbf{G}_3\cdot \mathbf{r}}
|
n,y_0,j,k_y\rangle=&
e^{-iGy_0}
e^{-iG\Delta j}
\delta_{y_0',y_0}
\delta_{j',j}
\int_{-\infty}^\infty d\bar{y}
e^{-iG\bar{y}}
\psi_{n'}
\left(\frac{\bar{y}}{l_B}\right)
\psi_n
\left(\frac{\bar{y}}{l_B}\right)
\equiv\hat{\mathcal{T}}_{j}^{(3)}.
\end{align}
The integrals over $y$ in the equations above are usually expressed in terms of generalized Laguerre polynomials to ease the numerical procedure\cite{Bistritzer_Moire_2011, Hejazi_Landau_2019} as
\begin{align}
\int dy e^{iG_yy}
\psi_{n'}
\left(
\frac{y}{l_B}-\frac{G_xl_B}{2}
\right)
\psi_{n }
\left(
\frac{y}{l_B}+\frac{G_xl_B}{2}
\right)=
\left\{
\begin{matrix}
    \left[
    \frac{(-G_x+iG_y)l_B}{\sqrt{2}}
    \right]^{n'-n}
    \sqrt{\frac{n'!}{n!}}
    \mathcal{L}_{n}^{n'-n}
    \left(
    \frac{G^2l_B^2}{2}
    \right)
    e^{-\frac{G^2l_B^2}{4}}\quad n'\geq n
    \\
    \left[
    \frac{(G_x+iG_y)l_B}{\sqrt{2}}
    \right]^{n-n'}
    \sqrt{\frac{n!}{n'!}}
    \mathcal{L}_{n'}^{n-n'}  
    \left(
    \frac{G^2l_B^2}{2}
    \right)
    e^{-\frac{G^2l_B^2}{4}},\quad n'<n
\end{matrix}
\right.
\end{align}
The matrix elements above allows us to compute the magnetic band structure numerically, setting the maximum the number of Landau levels $N_c$ that ensures convergence in the energy range of interest. Using the basis of magnetic Bloch functions
\begin{align}
\Psi_{\boldsymbol{k}}=\left(
    \begin{matrix}
        |0,y_0,j=0,k_y\rangle\\
        |1,y_0,j=0,k_y\rangle\\
        \vdots\\
        |N_c,y_0,j=0,k_y\rangle\\
        |0,y_0,j=1,k_y\rangle\\
        \vdots\\
        |N_c,y_0,j=q-1,k_y\rangle
    \end{matrix}
    \right),
\end{align}
the matrix form of the Hamiltonian, diagonal in the magnetic momentum $k_x=y_0/l_B^2$ and $k_y$, reads
\begin{align}
    \mathcal{H}=
    \left(
    \begin{matrix}
        \hat{E}_{n}^\beta+
        \frac{\gamma}{2}
        \left(
        \hat{\mathcal{T}}_{j=0}^{(3)}+\hat{\mathcal{T}}_{j=0}^{(3)\dagger}
        \right)&
        \frac{\gamma}{2}
        \left(
        \hat{\mathcal{T}}_{j=0}^{(1)\dagger}+
        \hat{\mathcal{T}}_{j=1}^{(2)}
        \right)&0&\cdots&\frac{\gamma}{2}
        \left(
        \hat{\mathcal{T}}_{j=q-1}^{(1)}+\hat{\mathcal{T}}_{j=0}^{(2)\dagger}
        \right)\\[1.5em]
        \frac{\gamma}{2}
        \left(
        \hat{\mathcal{T}}_{j=0}^{(1)}+        
        \hat{\mathcal{T}}_{j=1}^{(2)\dagger}
        \right)&
        \hat{E}_{n}^\beta+
        \frac{\gamma}{2}
        \left(    \hat{\mathcal{T}}_{j=1}^{(3)}+\hat{\mathcal{T}}_{j=1}^{(3)\dagger}
        \right)&
        \frac{\gamma}{2}
        \left(
        \hat{\mathcal{T}}_{j=1}^{(1)\dagger}+\hat{\mathcal{T}}_{j=2}^{(2)}
        \right)&\cdots&0\\[1.5em]
        0&
        \frac{\gamma}{2}
        \left(
        {\hat{\mathcal{T}}}_{j=1}^{(1)}+        
        \hat{\mathcal{T}}_{j=2}^{(2)\dagger}
        \right)&
        \ddots
        &\ddots&0\\[1.5em]
        \vdots&\vdots&\ddots&\ddots&\vdots\\[1.5em]
        \frac{\gamma}{2}
        \left(
        \mathcal{T}_{j=q-1}^{(1)\dagger}+
        \mathcal{T}_{j=0}^{(2)}
        \right)&0&0&\cdots&
        \hat{E}_{n}^\beta+
        \frac{\gamma}{2}
        \left(
        \mathcal{T}
        _{j=q-1}^{(3)}+
        \mathcal{T}_{j=q-1}^{(3)\dagger}
        \right)
    \end{matrix}
    \right),
\end{align}
where $\hat{E}_n^{\beta}=E_n\otimes\mathbb{1}_j$ ($\mathbb{1}_j$ is the unit matrix of dimension $j$). The Hofstadter's butterfly is obtained by plotting the bandwidth of all bands as a function of magnetic field.

\begin{figure}
\begin{center}
\includegraphics[width=1\columnwidth]{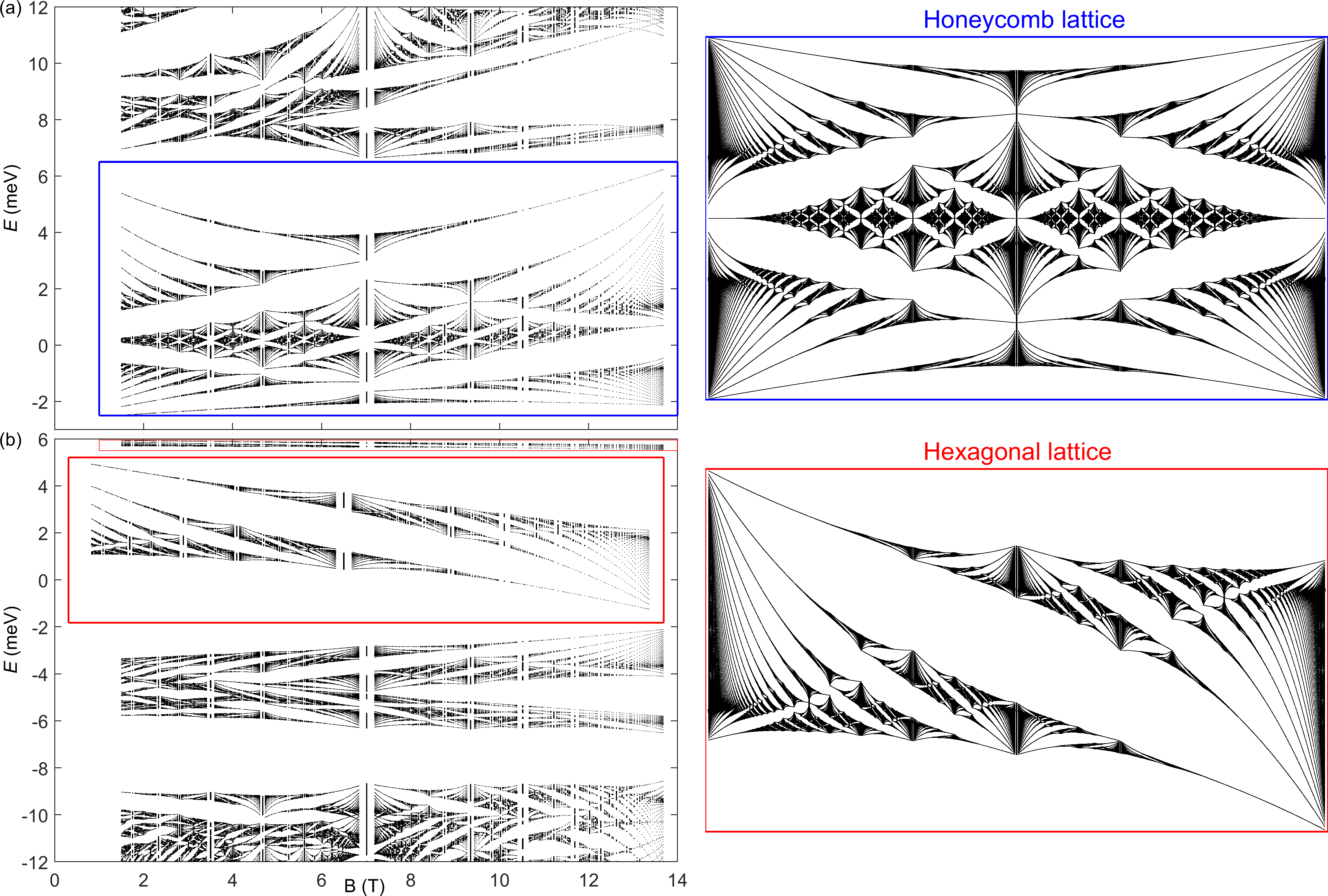}
\caption{ (a) Comparison between the electronic spectrum in the conduction side of \ch{MoS2} on a hexagonal superlattice potential, with $R_m=18.4$ nm and $\gamma=6$ meV, and Hofstadter's butterfly of graphene. Within the energy window spanning the first two bands, both spectra share identical features. (b) Electronic spectrum in the valence side of \ch{MoS2} on a hexagonal superlattice potential. The first group of bands are highlighted to visualize the Hofstadter's butterfly for a hexagonal lattice with a negative coupling constant, shown on the right. \label{Fig:Butterflies}}
\end{center}
\end{figure}

As stated in the main text, the Hofstadter's spectrum for electrons (holes) closest to the charge neutrality point inherits the symmetry of a honeycomb (hexagonal) lattice. Here, to further illustrate this feature, we plot in Fig. \ref{Fig:Butterflies} the Hoftadter's spectrta of \ch{MoS2} under a hexagonal superlattice following the above-described method with $R_m=18.45$ nm and $\gamma=6$ meV, below and above the charge neutrality point in panels (a) and (b), and to compare, we also present the Hofstadter's butterfly for a honeycomb and a hexagonal lattice, respectively. They were obtained using a tight binding model including a Peierls phase, for values of the magnetic fields that allows for a commensurate unit cell \cite{Peierls_Zur_1933,Rhim_Self-similar_2012, Li_Tight-binding_2011, Oh_Energy_2000}. The similarity of these spectra reassures inherently different symmetries of our system for electrons and holes. 